%% file: rootarx.tex
\documentclass[11pt,a4paper]{article}
\usepackage{a4wide}

\input clmacros1

\newcommand{\coli}{CoLi}
\newcommand{\jm}{Japaridze machines}

\begin{document}

\begin{center}
  {\Large {\bf  What is an Algorithm?: a Modern View}}
\\[20pt] 
{\bf Keehang Kwon }\\
Dept. of Computer Engineering, DongA University \\
% Busan 604-714, Korea\\
%051-200-7784 \\
khkwon@dau.ac.kr
\end{center}

Although algorithm is one of the central subjects, there have been
little common understandings of what an algorithm is.  For example,
Gurevich\cite{Gure11} view
 algorithms as abstract state machines, while others view  algorithms
 as recursors.  We promote a third view:   it is  a  combination to these two disparate views.
 
 This  approach -- based on computability logic\cite{Jap03} -- describes an algorithm as $A(I,O)$
 where $I$ is a set of input services and $O$ an output service.  It leads to the following modern definition: \\
 
 {\it An algorithm $A$ is a (tree of) sequence of legal moves  for 
 providing $O$ using $I$. } \\
 
  \noindent In the above, $A$ is written in an imperative language/abstract state machine and $I,O$ are written 
  in recursors/logical specifications.

%% </local definitions here>

\section{Introduction}\label{sec:intro}

There has been a declarative  approach to  algorithms.
In this approach,  algorithms are expressed
using other algorithms.
This approach  includes  recursive algorithms,
logical algorithms and computability logical(CoL)
algorithms\cite{Jap03,Jap08}.  Among these,
CoL algorithms is the most expressive of all.

Unfortunately, executing declarative algorithms are often slow due to high
nondeterminism in finding proofs. That is, it is often inefficient to automatically generate
proof steps. For this reason, it is often useful for the
programmer to specify proof scripts.
In other words, {\it combining imperative algorithms and declarative ones}
is often needed.

This paper proposes to use \coli\ as a new algorithm language. The distinguishing feature
of \coli\ over CoL is that now the programmer is allowed to provide
imperative features. 

\input intro2

\input prelim

\input mfor

\section{CoLi Algorithms}\label{sec:logic}

 A CoLi algorithm  is of the form

\begin{center}
 \[ A(I,O) \]
 \end{center}
 
 \noindent where 
 $I = \{ I_1,\ldots,I_n \}$ is a set of input services and $O$ is an output service, both written in CoL.
  $A$ is a winning strategy, i.e.,  a tree of runs\footnote{ A run is a sequence of moves.} written in pseudocode. A run 
is a sequence of $S_1;S_2;\ldots$ where $S$ is of several kinds. We list some of them below.

\begin{enumerate}

%\item $l.replicate$ replicates a subformula  located at $l$. For example, $l.replica$ replicates $p(a)$ at location
%$l$ to $p(a) \mlc p(a)$.

\item $l.read(N)$ replaces a subformula  $\ada x F$ at location $l$ by $F(X/x)$ where $X$ is a new variable.
 For example, suppose $\ada x p(x)$ is at $l$. $l.read(N)$ replaces it by $p(X)$
 and stores $X$ in the global variable $N$. 

\item $l.write$ replaces $\ade x F$  at $l$ by $F(W/x)$ where $W$ is a new global  variable.
For example, suppose $\ade x p(x)$ is at $l$. $l.write$ replaces $\ade x p(x)$ by $p(W)$.
 The value of $W$ is unknown and will later be determined via the unification process\footnote{Global variables are
 different from variables. The unification process only deals with global variables.}. This technique is well-known
 in the logic programming community.

\item  $choose(l_1:r_1,\ldots,l_n:r_n)$ limits the proof search space to the given $n$ choices,  where
each $l_i$ denotes a location of some subformula $H$ and $r_i$ is a list of  rule candidates 
to apply to $H$. We often omit $r_i$ when it is obvious.

\item $schoose(l_1:r_1,\ldots,l_n:r_n)$ is identical to $choose(l_1:r_1,\ldots,l_n:r_n)$ with the difference
 that $l_1:r_1$ has the highest priority and $l_n:r_n$ has the lowest.

\item Conditional statements include the if-then-else.
 Iterational statements include the for-loop.

\item $prove$ tries to extract a winning strategy $S$ from the current configuration. $execute$ executes $S$.

\end{enumerate}

An internal algorithm $A$ typically is very complex and challenging.  It requires two stages:

\begin{itemize}

\item The stage 1 extracts a winning strategy $S$ from the given $(I,O)$.
      $S$ is typically  nondeterministic and has the form of {\it a tree of runs} due to interactive services.
      This stage is very difficult due to the complexity of the proof procedure.

\item The stage 2 executes a branch of $S$ which is obtained by interacting with the user. This stage is easy.

\end{itemize}

In most cases, stage 1 can be automated by the machine. Unfortunately, there are cases when $S$ is
difficult to extract by the machine but the programmer knows $S$. In such cases, \coli\ would be 
useful. In writing \coli\ algorithms, it would be painful for the programmer to write $A$ from scratch.
Instead, $A$ is typically written by the programmer in a minimal way, i.e., in the style of proof scripts.
The rest will then be automatically generated by the machine.

 \section{Examples }\label{sec:modules}
 
As an  example, we present the factorial
 algorithm to help understand this notion.  The factorial algorithm can be
defined using  two input services $/c,/d$ whose tasks are 
 described below:
where  the recurrence action is preceded with $\pst$. \\ \\
%We assume that $fact$ is a constant of type $int \ra int \ra o$.

\noindent  $/c = fact(0,1)$\\
  $/d =  \pst\ \ada x\ada y(fact(x,y) \mli\ fact(x+1,xy+y))$.\\
%  $/e = !/c$  \% $the adress /e$ stores a copy of $fact(0,1)$\\
%  $/g = !/d$  \\
 %$/fac = /c \mlc /d$\\
%  $/fac1 = /e \mlc /g$\\
  $/query =  \ada y \ade z fact(y,z)$ \% read y and compute z (which is y! ) \\

\noindent  Suppose computation tries to solve the $/query$ with respect to $/c,/d$. Observe that
 $/query$ is a logical consequence of $/c$ and $/d$ and proving this fact  requires mathematical induction.
Extracting a winning strategy for this problem is quite difficult and nontrivial. Unlike the machine, however,
a winning strategy is obvious to human and is 
the following: read $y$ in the /query, make $y$ copies of $/d$, instantiate $x,y$ in 
each copy of $/d$ and then compute $z$ which is a logical consequence of these knowledgebase\footnote{
In fact, making a copy of $/d$ occurs on demand, i.e., when the corresponding $i$th component is not available yet.}.
The corresponding proof script $Fact$ is shown below:
 \\ \\

\noindent Algorithm $Fact(\{ /c,/d \}, /query)$ \% An algorithm for computing factorial. \\
/query.read(n); \\
%\noindent for i=1 to n; \\
%\hspace{3em} /d:replicate;  \% make 100 copies of /d \\
\noindent for i=1 to n;\\
\hspace{3em} /d.i.write;   \%  process $\ada x$ in the $i$th copy of /d. \\
\hspace{3em} /d.i.write;     \%  process $\ada y$ in the $i$th copy of /d. \\
endfor;\\
/query.write; \%  process $\ade z$ in the query \\ 
execute; \% invoke the unification procedure \\

In the above, note that $\ada y$ in the $/query$ is the major obstacle in extracting a winning strategy $S$.
Once $\ada y$ has been processed,  extracting the rest of $S$ poses no problem. Therefore, $Fact$ can be
simplified to the following: \\ \\

\noindent Algorithm $Fact(\{ /c,/d \}, /query)$ \% A shortened algorithm for computing factorial. \\
/query.read(n); \\
prove; \%  extract a winning strategy from the current configuration. \\ 
execute; \% invoke the unification procedure. \\

Proof scripts are also useful for dealing with semidecidable problems. For example, consider the following:

\[ /q = \pst \ade x p(x) \mld q(a). \]

\noindent The above formula is invalid.  Unfortunately, a typical proof procedure does not terminate for this formula,
as it repeatedly replicates $\pst \ade x p(x)$. We can avoid this unpleasant situation by providing the
following proof script which tells the machine {\it not} to replicate it. \\

\noindent  /q.1:$\ade$-rule;  \% replicate is disallowed in $\pst \ade x p(x)$.\\

\noindent In the beginning, the machine first chooses a term $t$ for $\ade x$ and obtains
$p(t) \mld q(a)$. Then  the proof procedure terminates with the failure.

\section{Conclusion}\label{sec:conc}

Our intention is to raise awareness of the CoLi algorithms as a new tool for 
expressing algorithms.  We believe it is a tool of real value.
Another interest is in designing a more flexible proof script language.

\bibliographystyle{ieicetr}% bib style

\end{document}

%% file: clmacros1.tex
\newcommand{\oo}{\bot}            %opponent, the corresponding truth value and label
\newcommand{\pp}{\top}            %proponent, the corresponding truth value and label
\newcommand{\pst}{\mbox{\raisebox{-0.01cm}{\scriptsize $\wedge$}\hspace{-4pt}\raisebox{0.16cm}{\tiny $\mid$}\hspace{2pt}}}
\newcommand{\gneg}{\neg}                  %game negation
\newcommand{\mli}{\rightarrow}                     %strong reduction
\newcommand{\mld}{\vee}    %multiplicative disjunction
\newcommand{\mlc}{\wedge}  %multiplicative conjunction
\newcommand{\ade}{\mbox{\Large $\sqcup$}}      %additive existential quantifier
\newcommand{\ada}{\mbox{\Large $\sqcap$}}      %additive universal quantifier
\newcommand{\add}{\sqcup}                      %additive disjunction
\newcommand{\adc}{\sqcap}                      %additive conjunction

\newcommand{\tlg}{\bot}               %classical \bot; trivially lost elementary game
\newcommand{\twg}{\top}               %classical \top; trivially won elementary game

\newtheorem{theoremm}{Theorem}[section]
\newtheorem{conditionss}{Condition}[section]

\newtheorem{definitionn}[theoremm]{Definition}
\newtheorem{lemmaa}[theoremm]{Lemma}
\newtheorem{notationn}[theoremm]{Notation}
\newtheorem{propositionn}[theoremm]{Proposition}
\newtheorem{conventionn}[theoremm]{Convention}
\newtheorem{examplee}[theoremm]{Example}
\newtheorem{remarkk}[theoremm]{Remark}
\newtheorem{factt}[theoremm]{Fact}
\newtheorem{exercisee}[theoremm]{Exercise}
\newtheorem{questionn}[theoremm]{Open Problem}
\newtheorem{conjecturee}[theoremm]{Conjecture}

%% file: intro2.tex
\section{Turing machines or Japaridze machines?}

The class of Turing machines (TMs) has been a standard model of computation.
It describes an  algorithm in the standard form of 

\[ A(i,o) \]
\noindent  where $i,o$ is an input/output  value, and $A$ is an internal imperative algorithm which maps $i$ to $o$.
 Thus TMs focuses on input-output mappings. 
 
 Japaridze\cite{Japi,Japfi,Japtow} proposed a new computing model which we call {\it \jm}.   
 It is a TM which focuses on its exchanging services(input services and output services).  That is, it  describes an  algorithm with respect to 
their exchanging services.  To be specific, it describes an algorithm in the form of 

\[ A(I,O) \]
\noindent where   $I$ is the set of input services and $O$ is the output service and $A$ is an internal imperative algorithm to accomplish $O$ using $I$.   
For example, consider a task "computes 3!".  While the conventional TM would produce 6, \jm\ produce
a service/knowledge which is $fact(3,6)$.  
We call this   approach  CoLi algorithms. Despite of its several advantages, 
it is quite unfortunate that  CoLi algorithms have been largely ignored by
academia and industry.
  
  We compare these two models. First,  TM preserve only the input/output behavior of a function.
There is more to an algorithm than the function it computes. \jm\  provides $services$ rather than function outputs.
The notion of services is a big concept which includes knowledge, interactive services, other complex services.

Second,  the single, low abstraction level of the Turing machine inhibits its ability to describe algorithms concisely.
 The author of   \jm\  can choose an arbitrary set of input services and therefore has flexibility in choosing the level of
abstraction. 

Third, given $I$ and $O$,   $A$ can automatically be generated by \jm. 
We call the  description $(I,O)$ {\it CoL algorithms}.  

Finally, it is easier to extend \jm\  to distributed computation.
A distributed \jm\ (also know as computability-logic web\cite{Kee2020}) is a  
set of  \jm\ providing services to one another.  It is a promising model for distributed computing with several attractive
features such as local name space and service migrations.

Turing machines lead to the development of assembly languages and C.
We now consider how \jm\  could be useful in a new language development.
In the $A(I,O)$  above, note that  $A$ is written in  imperative languages whereas $I,O$ are written in logic languages
in CoL.

 This  lead us to a next-generation imperative language where  imperative languages are used as implementation
 languages and logic languages are  
 used as {\it  specification}.  This new language is thus closely related to the deep specification approach to software.
 An example may be Python with {\it logical  specification of input/output services}. That is,

\begin{itemize}

\item  Turing machines $\Rightarrow$ Assembly languages, C, Python, $\ldots$

\item \jm $\Rightarrow$  Assembly+deep specification, C+deep specification, $\ldots$

\end{itemize}

We now consider when we need CoLi algorithms.
CoL is a complex language with  a huge yet of useful operations. The design of 
\jm\ aims at automatically generating an internal algorithm/strategy  from given input and out services. 
This approach has been 
successful for various fragments of CoL. Yet, implementing the full CoL is a totally different story: it seems a daunting, almost 
impossible task
due to its huge complexity. Accordingly, we have no other choice but to rely on CoLi algorithms to utilize the full CoL.

%% file: prelim.tex
\section{Preliminaries}\label{s2}

In this section a  brief overview of CoL is given. 

There are two players: the machine $\pp$ and the environment $\oo$.

There are two sorts of atoms: {\em elementary} atoms $p$, $q$, \ldots to represent elementary games, and {\em general atoms} $P$, $Q$, \ldots to represent any, not-necessarily-elementary, games. 

\begin{description}
\item[Constant elementary games]  $\twg$ is always a true proposition, and $\tlg$ is always a false proposition.

\item[Negation]
 $\gneg$ is a role-switch operation: For example, $\gneg (0=1)$ is true,
while $(0=1)$ is false.

\item[Choice operations]
The choice group of operations:  $\adc$, $\add$, $\ada$ and $\ade$ are defined below.

$\ada xA(x)$ is the game where, in the initial position, only $\oo$ has a legal move which consists in 
choosing a value for $x$. After $\oo$ makes a move $c\in\{0,1,\ldots\}$, 
the game continues as $A(c)$. 
$A\adc B$ is similar, 
only here the choice is  made between ``left"  and ``right". $\ade$ and $\add$ are 
symmetric to $\ada$ and $\adc$, with
the  difference that now it is $\pp$ who makes an initial move.

\item[Parallel operations]
Playing $A_1\mlc\ldots\mlc A_n$ means playing the $n$ games concurrently.  In order to win,  $\pp$ needs to win in each of $n$ games. Playing  $A_1\mld\ldots\mld A_n$ also means playing the $n$ games concurrently.  In order to win,  $\pp$ needs to win  one of the games. To indicate that a given move is made in the $i$th component, the player should prefix it with the string ``$i.$".  
The operations $\pst A$ means an infinite parallel
game $A\mlc\ldots\mlc A\mlc\ldots$.
 To indicate that a given move is made in the $i (i>1)$th component, we assume the player should 
 first replicate $A$ and then prefix it with the string ``$i.$".

\item[Reduction]
 $A\mli B$ is defined  by $\gneg A\mld B$.
Intuitively, $A\mli B$ is the problem of reducing $B$ ({\em consequent}) to $A$ ({\em antecedent}).  

\end{description}

%% file: mfor.tex
\section{Introducing Directories }\label{sec:intro}

Logical formulas  are inadequate for locating subformulas. 
Our approach to achieving this effect is through the use of 
{\it directories}. For example, consider the following  directory
definition.

\[ /m = p(a) \]

\noindent where $/m$ is a directory name and $p(a)$ is a formula.
In this case, we call $p(a)$ its ``content''.
Alternatively, we can view $/m$ as an agent and $p(a)$ as its
knowledgebase.

Our directory system is very flexible and is designed
to represent both formulas and cirquents.
For example, $/n = !/m \land !/m$ represents that the directory $/n$
contains $p(a) \land p(a)$.
 Here $!/m$ is intended to read as ``a copy of the
 content of $/m$. In contrast, $/o = /m \land /m$ represents that
 $/o$ contains a cirquent
 $p(a) \land p(a)$ where two $p(a)$s in $/o$ and $p(a)$ in $/m$
 are $shared$.

% max(L)
% m = 1,2,3,4 =>  m,m

As another example, consider the following  recursive directory
definition.

\[ /m(0) = q \]
\[ /m(s(X)) = p \land !/m(X) \]

\noindent Given this definition, $p \land (p \land (p \land q)))$ can be represented simply as $ /m(s(s(s(0))))$.  We assume in the above that
$s$ is the number-successor function.

Thus, we propose the notion of {\it directorized}
formulas.  They are  formulas enhanced with 
directories. These formulas are better-suited to structuring large
formulas such as pigeonhole
principle formulas.
It is interesting to note that directories also play the role of global
variables
in
imperative languages and much more.